\documentclass[sigconf,authorversion,nonacm]{acmart}

\AtBeginDocument{}
\setcopyright{acmcopyright}
\copyrightyear{20XX}
\acmYear{20XX}
\acmDOI{XXXXXXX.XXXXXXX}

\acmConference[Conference acronym 'XX]{Make sure to enter the correct conference title from your rights confirmation email}{Month, days, 20XX}{City, State}
\acmPrice{15.00}
\acmISBN{978-1-4503-XXXX-X/18/06}

\usepackage{xcolor}
\usepackage{hyperref}
\usepackage{enumitem}
\usepackage{wrapfig}
\begin{document}

\newcommand{\DarkDiff}{DarkDiff}
\newcommand{\DWM}{DWM}
\newcommand{\TUDelft}{TUDelft}
\newcommand{\CFLW}{CFLW}
\newcommand{\DarkWebMonitor}{Dark Web Monitor}
\title{\DarkDiff{}: Explainable web page similarity of TOR onion sites}


\author{Pieter Hartel}
\authornote{The first author wrote the paper and all authors contributed to the analysis of the data}
\email{Pieter.Hartel@tudelft.nl}
\orcid{0000-0002-0411-0421}
\affiliation{%
\institution{Delft University of Technology}
\institution{CFLW Cyber Strategies B.V.}
\city{}
\country{The Netherlands}
}
\author{Eljo Haspels}
\email{Eljo.Haspels@cflw.com}
\orcid{0009-0004-3304-877X}
\author{Mark van Staalduinen}
\email{Mark.vanStaalduinen@cflw.com}
\orcid{0000-0001-9716-3815}
\author{Octavio Texeira}
\email{Octavio.Texeira@cflw.com}
\orcid{0009-0000-4253-1916}
\affiliation{%
\institution{CFLW Cyber Strategies B.V.}
\streetaddress{Franklinstraat 1A}
\state{}
\country{The Netherlands}
\postcode{2691 HB}
}

\renewcommand{\shortauthors}{Hartel et al.}

\begin{abstract}
In large-scale data analysis, near-duplicates are often a problem.
For example, with two near-duplicate phishing emails, a difference in the salutation (Mr versus Ms) is not essential, but whether it is bank A or B is important.
The state-of-the-art in near-duplicate detection is a black box approach (MinHash), so one only knows that emails are near-duplicates, but not why.
We present \DarkDiff{}, which can efficiently detect near-duplicates while providing the reason why there is a near-duplicate.
We have developed \DarkDiff{} to detect near-duplicates of homepages on the Darkweb.
\DarkDiff{} works well on those pages because they resemble the clear web of the past.
\end{abstract}





\maketitle

\section{Introduction}
Some web documents change frequently, and others rarely or never \cite{Lim2001}.
The dynamics of web pages on the clear web have been extensively researched \cite{Olston2010}.
Changes are small, and not all pages change.
Previous changes are a good predictor of new changes, and pages on certain top-level domains change more often \cite{Adar2009}, \cite{Cho2003}, \cite{Fetterly2004}, \cite{Ntoulas2004}.
Many changes are trivial, such as timestamps and visitor counts \cite{Douglis1998}, \cite{Henzinger2006}.
Therefore, a WWW crawler encounters many (near) duplicates \cite{Burda2019}.
Google defines near-duplicates as "documents identical in terms of content but different in a small portion of the document such as advertisements, counters and timestamps" \cite{Manku2007}.

An onion site is a website that is reachable only via the Tor network (See \url{https://www.torproject.org}).
The nature and frequency of changes to onion sites are probably different from clear-web sites.
For example, Search Engine Optimization (SEO) hardly plays a role for onion sites \cite{Gehl2021}.
There are search engines for onion sites, such as \url{https://ahmia.fi}, but there is yet to be a market for ranking onion sites.
Hence onion sites do not change as often, nor do they change in the same way as search-optimised websites for the clear web.
As another example, all sites on the clear web use JavaScript \cite{Yue2009}.
The general recommendation is that onion sites should not be visited with JavaScript enabled in the browser \cite{Nikiforakis2012}.
Onion sites often function without JavaScript, unlike the clear web.
Therefore changes to an onion site are usually changes to the HTML, whereas changes to a clear-web site also include changes to the JavaScript.

Crawlers of websites must be able to handle near-duplicates efficiently because near-duplicates often represent noise in the data \cite{Theobald2008}.
Near-duplicates arise when a crawler repeatedly indexes the same site but also when the crawler indexes a phishing site or a mirror.
There are efficient techniques for detecting near-duplicates, such as MinHash \cite{Broder1997} and SimHash \cite{Charikar2002}.
However, these black-box techniques do not indicate why documents are near-duplicates.
Most near-duplication detection algorithms represent documents as a set of words and calculate (an estimate of) the Jaccard index.
If the Jaccard index of two documents exceeds a chosen threshold value, it is assumed that the documents are near-duplicates.
The problem is that the Jaccard index does not indicate why that would be the case.
For example, which words are in the first but not the second document?
MinHash and Simhash are not transparent.
Our research question is: how to recognise a near-duplicate efficiently \emph{and} explain why it is a near-duplicate?

We discuss several ways to recognise near-duplicates in the Background section.
We first provide the intuition for the solution we explore in this article.
Suppose a crawler visits the same web page at two different times, and the two versions of the web page are v1 = "Lorem 2023-04-05 ipsum" and v2 = "Lorem 2023-04-06 ipsum".
The difference between v1 and v2 is that v1 contains the date one day earlier than v2.
We define the diff between two documents, v1 and v2, as the smallest set of deleted or inserted words that converts v1 to v2.
Then the diff of v1 and v2 is d1\_2 = "Lorem {\color{red}2023-04-05}{\color{green}2023-04-06} ipsum".

Assume that the differences (i.e. red and green strings) are dates.
Then we can replace the two differences with a reserved word, for example, {\color{blue}date}.
Replacing differences with reserved words creates what we call an annotated template.
In the example, we obtain: t1 = t2 = "Lorem {\color{blue}date} ipsum".
Since t1 = t2, we conclude that v1 and v2 are near-duplicates because they differ in dates.

Finally, We can calculate the hash of the annotated templates to quickly determine whether pages have the same annotated template and are, therefore, near-duplicates.
The hashes of the annotated templates can be stored in a database to discover near-duplicates efficiently.
Table \ref{tab:summary} summarises the \DarkDiff{} process.

\begin{table}
\caption{\DarkDiff{} applied to two near duplicates equal except for the date. The arrows indicate that a diff has two inputs and two outputs}
\label{tab:summary}
\begin{tabular}{lrlll}
\toprule
	&Page version		&\DarkDiff{}				&	&Annotated template \\
\midrule
v1	&\multicolumn{2}{l}{Lorem 2022-04-05 ipsum}	$\nearrow$	&t1	&Lorem {\color{blue}date} ipsum \\
	&	$\searrow$	&Lorem {\color{red}2022-04-05}		&	& \\
	&	$\nearrow$	&{\color{green}2022-04-06} ipsum	&	& \\
v2	&\multicolumn{2}{l}{Lorem 2022-04-06 ipsum}	$\searrow$	&t2	&Lorem {\color{blue}date} ipsum \\
\bottomrule
\end{tabular}
\end{table}

Reserved words like {\color{blue}date} form the "explanation of the similarity of the near-duplicates".
Suppose we have two different web page versions: v1 = "Lorem 2023-04-05 ipsum" and v3 = "Lorem 1 BTC ipsum", for which
\DarkDiff{} will generate two different annotated templates: "Loren {\color{blue}date} ipsum" and "Loren {\color{blue}price} ipsum".
The conclusion is that the two pages are not near-duplicates; the explanation is that one contains a price and the other a date.

On the clear web, many differences cannot be interpreted with a regular expression, for example, different advertisements woven into the text of a press release \cite{Theobald2008}.
We hypothesise that most changes between different page versions on onion sites are relatively simple compared to the clear-web sites.

Our contribution is threefold:
\begin{itemize}[noitemsep,topsep=0pt]
\item We present \DarkDiff{} as a tool for explainable similarity and compare the performance with the Jaccard index.
\item We analyse random samples from a large dataset of onion sites for near-duplicates of the homepage.
We thus show that \DarkDiff{} is scalable.
\item We show how \DarkDiff{} uncovers interesting facts in three case studies.
\end{itemize}

\section{Background}
A web page may undergo many changes during its lifecylce, for example as a result of maintenance, updates, and search engine optimisation.
We discuss studies that have examined differences in clear web pages over time.
However, we will start the background section with a summary of the differences between onion sites and clear-web sites because the studies of clear-web sites do not necessarily apply to onion sites.

\subsection{Onion sites and branding}
People who use onion sites care about their privacy, such as whistleblowers, but also criminals and terrorists.
There are all kinds of onion sites for legitimate purposes but also for illegitimate ones \cite{Jardine2018}.
Website names are essential to an organisation's branding on the clear web.
For example, most people will recognise google.com and google.co.uk as websites of Google Inc.
An onion site does not have a meaningful name, as it is based on a public key.
This makes it difficult to use the name of an onion site for branding purposes.
The randomness also makes it difficult for users to identify phishing sites \cite{Winter2018}.
Some onion sites use subdomains for branding purposes.
Onion sites are copied on a large scale \cite{Yoon2019}.
The idea is that the more subdomains there are, the more visitors will be lured to the onion site.
As a result, there are many near-duplicates of onion sites, which is a challenge for software that tries to show relevant content, like search engines.

\subsection{Homepages, access barriers, and redirections}
For most onion sites, the homepage is the business card of the site because it is the first page visitors encounter.
The homepage usually also shows what the purpose of the site is.
The homepage of each onion site is freely accessible, but sometimes the homepage is a login page, a page with a captcha, a page with a timer, or a combination.
All of these require either human intervention or machine learning \cite{Dionysiou2019} to bypass the access barrier, which is beyond the scope of this paper.

The homepage of an onion site is sometimes a redirect to another page.
There are three ways a server can redirect the browser \cite{Chellapilla2007}:
(1) When the server returns a redirect in an HTTP request to the browser, the browser will immediately use the new URL in the HTTP response to load the desired page.
Because an HTTP redirect takes place before a web page is loaded, it is not considered here.
(2) The web page includes the following meta element:
"<meta http-equiv='refresh' content='0; url=https://target.onion/'>".
A browser will not show such a "meta-refresh" page but will immediately try to load the target page (indicated by the URL attribute).
(3) The web page contains a JavaScript program that automatically executes the redirect, and also, for browsers with JavaScript disabled, a link that the user can click on, for example:
"<script> window.location.replace('http://target.onion/'); </script>",
and "<a href='http://target.onion/'>Wait a second...</a>".
JavaScript is Turing complete, so it is only possible to determine what the code will do by executing it.
Therefore, we assume we are dealing with a redirect if we encounter an anchor element with a text such as "Wait a second...".
Also, in case (3), the browser will not show the page with the script but the target page.

On the clear web, redirection pages are SPAM because they provide no content \cite{Ntoulas2006}.
Our research looks at homepages, avoids redirects as much as possible, and stops at access barriers.

\subsection{Frequency and nature of changes}
Cho and Garcia-Molina \cite{Cho2000} define the visible life span as the number of days the crawler could find a page and the
average change interval as the number of days between observed changes.
A U curve describes the relationship between the fraction of domains and the average change interval.
Thus there are two groups of pages: those that change often and those that change rarely.
The relationship between the fraction of domains and the visible life span is exponential, and the lifespan of pages follows a Poisson process.

Web pages change more frequently now than in the early days of the web \cite{Olston2010}.
There are now more important differences between near-duplicates on the clear web than before, for example adverts, and syndicated news that did not exists in the early days of the web
We have found that changes to current onion sites resemble the changes to clear-web sites of the past.
See Appendix \ref{app:past} for the details.

\subsection{Near duplicates}
Near duplicates can be found efficiently with locality-sensitive hashing (LSH).
Locality-sensitive hashing is a technique to map documents of arbitrary size to a relatively small fingerprint.
If two documents have a lot in common, the fingerprints of the two documents will not differ much from each other.
Well-known LSH algorithms are MinHash \cite{Broder1997}, and SimHash \cite{Charikar2002}.

For the application of LSH to HTML pages, all markup and extra white space are usually removed, and lowercase and uppercase are not distinguished \cite{Lim2001}.
Information that is removed cannot play a role in explaining why two documents are near-duplicates.
Some authors \cite{Fetterly2004} assume that the text of a link has meaning so that the link itself is superfluous.
However, an HTML element such as "<a href=...>click here</a>" does not always satisfy the assumption of Fetterly et al. since "click here" contains no information.
Henzinger \cite{Henzinger2006} also omits most of the markup but makes an exception for the HREF attributes of an anchor element.
She applies the algorithms of Broder, Charikar, and a combination of the two to a vast dataset of 1.6B web pages.
The combination works best.

Zhang et al. \cite{Zhang2016} use the Normalised Compression Distance to compare webpages where the <script>, <style>, and <iframe> elements have been replaced by whitespace.

Boldi et al. \cite{Boldi2018} describe an open-source crawler for the web.
They strip the HTML of markup, numbers and dates so that even more information is lost.
This heuristic allows grouping pages that differ in, for example, the number of visitors.
\DarkDiff{} also drops information, but we remember what type of information it was and where it occurred.

Froebe et al. \cite{Froebe2021a} study several methods of extracting content from a web page, none of which take markup into account.
We include the entire HTML, including markup, in the analysis because we want to be able to explain the why of the near-duplicates.

\subsection{Diffing}
Diffing is a technique for comparing linear lists (the lines of two documents).
However, an HTML document has a hierarchical structure, not a linear one.

Mikhaiel et al. \cite{Mikhaiel2005} discuss a system where the inserts and deletes of the diff respect the tree structure of the HTML.

Douglis et al. \cite{Douglis1998} describe a tool to create diffs from HTML pages.
Future work proposes that mundane differences such as "there were 1234 visitors to this site" should be filtered out.

Shannon et al. \cite{Shannon2010} show how diffing can visualise the history of a document.

Borgolte et al. \cite{Borgolte2014} use a fuzzy tree difference algorithm to track the changes to a webpage.

Layton et al. \cite{Layton2009} cluster web pages using standard NLP techniques and compare the results with those same techniques applied to the diff of the pages rather than the pages themselves.
So the work of Layton et al. is a form of \DarkDiff{}.
The difference between the two methods of Layton et al. is slight, but the benefit lies in the reduction of the dataset by 23\%.
Our research uses the diffs differently to recognise the changes with regular expressions.

\subsection{JavaScript and redictions}
Web pages contain markup, text, and code.
The code contains the most variation.
Therefore, we expect web pages with JavaScript to be more of a problem for near-duplicate detection than web pages without code.

A study of 6,805 popular domains, according to Alexa.com, found that in 2008 96.6\% of sites already used JavaScript.
66.4\% of the sites were more vulnerable than those without JavaScript \cite{Yue2009}.
One of the reasons for those vulnerabilities is that a website can load JavaScript libraries from other sites.
Those third-party libraries can be compromised or contain vulnerabilities.

JavaScript is also a problem for archiving.
Goel et al. \cite{Goel2022} show that the quality of web archives suffers from using JavaScript.
They propose a method to overcome the problems.
We investigate to what extent JavaScript is detrimental to \DarkDiff{}.

\subsection{Bitcoin scams}
Bitcoins are an essential means of payment on the clear web, but even more so on onion sites, which creates opportunities for fraudsters.
Lee et al. \cite{Lee2019} found 10K Bitcoin addresses in 27M onion sites, 80\% of which can be associated with criminal activity.
Winter et al. \cite{Winter2016} report that some exit node owners fraudulently replace Bitcoin addresses with their own addresses.
The owner of the exit node then receives any payments instead of the victims of the fraud.
Badawi et al. \cite{Badawi2022} found 8,000 Bitcoin addresses on the clear web during a 16-month study, all of which promise a multiplication of the deposits.
The multiplication would be possible because of a bug in Bitcoin.
The "profit" per address is about US\$46.
Bitcoin scams are also popular on onion sites.
In one of the case studies, we show how \DarkDiff{} discovered Bitcoin fraud.

\subsection{Research Ethics}
In the \DarkDiff{} study, we only perform secondary analysis of existing data owned by \CFLW{} as well as data from the public Bitcoin blockchain.
All authors are employees of \CFLW{}. The secondary analysis took place exclusively on the servers of \CFLW{}. Therefore, the research falls outside the scope of the IRB of the \TUDelft{}. In Appendix \ref{app:secondary} we use the Menlo report \cite{Dittrich2012} as a guideline in the analysis of the ethical risks and mitigations brought about by the \DarkDiff{} research.

\section{Method}
We discuss \DarkDiff{} in detail, followed by the \DWM{} dataset and the experiments with \DarkDiff{} on the dataset.

\subsection{\DarkDiff{}}
Given a series of $n$ different versions of a web page v1, v2, v3, \dots, we calculate the $n$ annotated templates t1, t2, t3 etc. in six steps:

\paragraph{Step 1. Tokenise the web pages}
The HTML of the page is not parsed but divided into tokens using a regular expression (RE):
"((?:\textbackslash d\{1,4\}[./-])\{2\}\textbackslash d\{1,4\}|
\textbackslash d+.\textbackslash d+|
\textbackslash w+|
\textbackslash s+|
\textbackslash W)".
The idea is not to break down the most common differences between versions but to treat them as a token.
Therefore, the first pattern in the RE matches the most common dates (e.g. "20/02/2023").
The second pattern "\textbackslash d+.\textbackslash d+" matches numbers with a decimal point often used for prices and exchange rates (e.g. "3.1415").
The third pattern "\textbackslash w+" matches a series of letters, numbers and underscores so that onion and Bitcoin addresses are treated as one (e.g. "silkroad6ownowfk"),
The fourth pattern "\textbackslash s+" matches white space.
The last pattern "\textbackslash W" matches everything that has not yet been matched, for example, "$<$".

\paragraph{Step 2. Use diff-match-patch to compute changes}
From $n$ versions, compute a series of $n-1$ diffs using Google's diff-match-patch (See \url{https://github.com/google/diff-match-patch}) on a token basis as follows:
d1\_2=diff(v1, v2); d2\_3=diff(v2, v3); etc.
Each di\_j is a list of textual deletes (in red), inserts (in green), and common texts (in black).
Table \ref{tab:no_alignment} gives an example of the calculation with the three versions v1, v2 and v3.
In the last column, we annotate the different templates.
Each diff creates two annotated templates, one due to the deletes and the other due to the inserts.
In the example, v2 creates two annotated templates t2 and t2', which are unequal.
We are going to address this problem with the alignment step 3 below.

\begin{table*}
\caption{Computing the annotated templates for three versions without alignment}
\label{tab:no_alignment}
\begin{tabular}{lrlrll}
\toprule
	&Page version			&	&\DarkDiff{}				&	&Annotated template \\
\midrule
v1	&Lorem 2022-04-05 ipsum 1.0 BTC	&	&			$\nearrow$	&t1	&Lorem {\color{blue}date} ipsum 1.0 BTC \\
	&		$\searrow$	&d1\_2	&Lorem {\color{red}2022-04-05}		&	& \\
	&		$\nearrow$	&	&{\color{green}2022-04-06} ipsum 1.0 BTC&	& \\
v2	&Lorem 2022-04-06 ipsum 1.0 BTC	&	&			$\searrow$	&t2	&Lorem {\color{blue}date} ipsum 1.0 BTC \\
	&				&	&			$\nearrow$	&$\neq$t2'
												&Lorem 2022-04-06 ipsum {\color{blue}price } BTC \\
	&		$\searrow$	&d2\_3	&Lorem 2022-04-06			&	& \\
	&		$\nearrow$	&	&ipsum {\color{red}1.0} {\color{green}1.5} BTC
											&	& \\
v3	&Lorem 2022-04-06 ipsum 1.5 BTC	&	&			$\searrow$	&t3	&Lorem 2022-04-06 ipsum {\color{blue}price} BTC \\
\bottomrule
\end{tabular}
\end{table*}

\paragraph{Step 3. Align the changes in chunks of diffs}
Not all near-duplicates need to have the same differences.
For example, Table \ref{tab:no_alignment} shows that v1 and v2 have different dates, but v2 and v3 have the same date.
Since there is no difference between the dates of v2 and v3, nothing can be annotated because the \DarkDiff{} regular expressions work on the differences.
One way to address the problem would be to apply diff-match-patch and regular expressions to v1 and v3.
That would solve the problem in the example because v1 and v3 have different dates and amounts.
However, the complexity would be quadratic in the number of versions.
Instead, we have chosen a sub-optimal, linear solution, which we call "alignment":
\emph{Where there are differences in an earlier or later diff, but where there is no difference in the current diff, the common text is broken up to create a "pseudo" difference.}

We indicate these pseudo differences in orange, see Table \ref{tab:perfect_alignment}.
For example, the text "ipsum 1.0 BTC" will be split into "ipsum", "{\color{orange}1.0}", and "BTC".
The alignment step splits common text, and no texts are merged.
It follows that the complexity of the step is linear in the number of diffs.
A unification algorithm would give an optimal result, but we prefer a linear algorithm for scalability.
The alignment has solved the problem of Table \ref{tab:no_alignment}: in Table \ref{tab:perfect_alignment}, we see that all annotated templates have become the same.

\begin{table*}
\caption{Perfectly annotated templates for three page versions with alignment}
\label{tab:perfect_alignment}
\begin{tabular}{lrlrll}
\toprule
	&Page version			&	&\DarkDiff{}			&	&Annotated template \\
\midrule
v1	&Lorem 2022-04-05 ipsum 1.0 BTC	&	&		$\nearrow$	&t1	&Lorem {\color{blue}date} ipsum {\color{blue}price} BTC \\
	&		$\searrow$	&d1\_2	&Lorem {\color{red}2022-04-05}	&	& \\
	&		$\nearrow$	&	&{\color{green}2022-04-06} ipsum
							{\color{orange}1.0} BTC	&	& \\
v2	&Lorem 2022-04-06 ipsum 1.0 BTC	&	&		$\searrow$	&t2	&Lorem {\color{blue}date} ipsum {\color{blue}price} BTC \\
	&				&	&		$\nearrow$	&=t2'	&Lorem {\color{blue}date} ipsum {\color{blue}price } BTC \\
	&		$\searrow$	&d2\_3	&Lorem {\color{orange}2022-04-06}
										&	& \\
	&		$\nearrow$	&	&ipsum {\color{red}1.0} {\color{green}1.5} BTC
										&	& \\
v3	&Lorem 2022-04-06 ipsum 1.5 BTC	&	&		$\searrow$	&t3	&Lorem {\color{blue}date} ipsum {\color{blue}price} BTC \\
\bottomrule
\end{tabular}
\end{table*}

The diff-match-patch step needs two inputs so that the next diff-match-patch step is independent of the previous one.
In Table \ref{tab:perfect_alignment}, d1\_2 and d2\_3 can be calculated independently.
However, the alignment step requires all diffs (here d1\_2 and d2\_3), introducing a dependency.
To make the alignment practical, we align a fixed number of diffs.
The more diffs the alignment step includes, the more the annotated templates will resemble each other.
However, there is also a limit to the number of diffs the alignment can include.
We will return to the number of diffs to align in Step 5.

\paragraph{Step 4. Annotate changes by regular expressions}
We apply regular expressions to the differences in a specific context to prevent inserts and deletes from being annotated incorrectly.
For example, the differences between the numbers in "ipsum {\color{red}1.0}{\color{green}1.5}BTC" are prices because there is a currency symbol immediately after the number.
We do not annotate a difference without the context of a known currency symbol to minimise the risk of incorrect annotations.
The smaller the risk, the more confidence we can have in the explainability of the changes and the more transparent the result.
The annotation itself consists of replacing the matched insert or delete with a reserved word, here {\color{blue}price}.
We will investigate to what extent the regular expressions correctly annotated an insert or delete with a random sample of deletes and inserts.

\paragraph{Step 5. Determine the chunk size}
In the example of Table \ref{tab:no_alignment}, there is no alignment; in other words, the chunk size is 0.
There are then two annotated templates.
In the example of Table \ref{tab:perfect_alignment}, two consecutive diffs are aligned, so the chunk size is 2.
The number of annotated templates drops down to 1.
We will investigate how the chunk size relates to the number of annotated templates.

\paragraph{Step 6. Store the hash of the annotated templates}
Store the hash of the annotated templates in a database.
Each hash can be looked up with an index to find all near-duplicates, modulo date, price, etc.

\subsection{Data set}
The crawlers of the \DarkWebMonitor{}(\DWM{}) have collected data from the Darkweb since 2013.
For online onion sites, one of the crawlers visits every 18 hours and retrieves a new version of the homepage.
For offline sites, the crawlers check the status every ten days.
The crawlers store the metadata of all retrieved pages in a database.
If the crawler revisits a page and finds that everything has stayed the same, it does not create a new version.

When the crawlers find a new onion site, that site is hand-tagged to several categories, the most important of which for this article are as follows:
{\bf Property offence} is an offence to obtain money, property or any other advantage, for example: Selling stolen credit cards.
{\bf Drugs offence} is possessing a controlled drug, such as the stocks of a Drugs online shop.
{\bf Child Sexual offence} is a sexual act prohibited by law, for example: Offering Child Sexual Abuse material (CSAM).
{\bf Violent offence} is the (threat of) use of harmful violence, for example: Hiring an assassin.
The tagging information allows us to compare the \DarkDiff{} performance for different types of onion sites.

\subsection{Procedure}
The analysis procedure consists of the following steps:

(1) We select all onion sites discovered by the \DarkWebMonitor{} in the period June 2020 until September 2022.

(2) We apply the following exclusion criteria to these onion sites:
(a) The site has no homepage.
(b) The homepage is an error page of the TOR proxy.
(c) We uniformly and randomly choose a certain percentage of the onion sites and exclude the rest.

(3) On the homepages of the selected site, we apply the following exclusion criteria:
(d) A version of the homepage of a site tagged as Child Sexual Offense has embedded images.
Possession of CSAM is illegal, so we will not consider CSAM pages with embedded images.
(e) A homepage version consists exclusively of redirection to another site, for example, with an immediate meta-refresh tag.
A reference does not provide original content, so we will not consider it.
(f) The crawlers found one version of a homepage, or as a result of (d) or (e), one version of the homepage remains.
\DarkDiff{} needs at least two homepage versions, so we will not consider singleton versions.
(g) We analyse the first 100 versions of the homepage.
This limitation means that we follow the changes in a homepage over at least 100 x 18 hours = 10 weeks, and at most the entire two year period.
Recently discovered onion sites and onion sites whose homepage stays mostly the same sometimes have fewer than 100 versions.

(4) We calculate the hash over all versions of all pages and generated annotated templates and report the percentage of near-duplicates found modulo date, prices etc.
We thus provide a metric for the success of \DarkDiff{}.
We also report how many onion sites share a common annotated template.

(5) The most critical parameter of \DarkDiff{} is the number of diffs included in the alignment step; we call this the chunk size.
We perform a sensitivity analysis on the chunk size by calculating near-duplicate percentages with these chunk sizes: 0, 2, 5, 10, 20, 50 and 100.

(6) We compare the Jaccard similarity with the Levenshtein similarity calculated based on the inserts and deletes from the diffs.

(7) \DarkDiff{} is a rule-based system that requires maintenance.
We test on a fresh dataset to what extent the rules need to be updated

(8) \DarkDiff{} focuses attention on the differences between homepages.
We examine differences in detail in three case studies and share our insights.

\section{Results}
From June, 2020, until September, 2022, the \DarkWebMonitor{} crawlers have discovered 2,505,904 onion sites with 23,537,693 homepage versions. 
The effect of the exclusion criteria is as follows:
(a) Of the sites discovered, 393,291 have a homepage. 
(b) 560 sites have an error page from a proxy as their homepage. 
(c) We uniformly and randomly select 10\% of the eligible sites, which amounts to 39,213 sites. 
(d) 498 versions of a homepage tagged as CSAM contain embedded images. 
(e) 18,002 homepage versions are immediate meta refreshes, and 2,148 are script refresh pages. 
(f) Criteria (d) and (e) leave 14,291 onion sites with at least two versions of the homepage. 
(g) We limit the number of homepage versions per onion site to 100. 
This leaves for analysis 443,194 versions of the homepage (an average of 31 per site). 

We provide the statistics for the main steps of \DarkDiff{}.

\paragraph{Step 2. Use diff-match-patch to compute changes}
Diff-match-patch generated 428,903 diffs from the 443,194 versions. 
Because the underlying algorithm is an optimisation, Google has built in the possibility to limit the calculation time.
We use the default of 1 second.
The total time required was 58,630s, on average 132 ms per diff (on an x86\_64 3.6 GHz UBUNTU system). 

\paragraph{Step 3. Align the changes in chunks of diffs}
Diff-match-patch generated 22,132,070 changes, each consisting of a delete, an insert, and common text. 
The alignment process sometimes splits the common text, resulting in 23,478,389 aligned changes. 
These numbers apply with chunk size ten.

\paragraph{Step 4. Annotate changes by regular expressions}
94.1\% of the aligned changes are annotated. 
The total time needed to annotate the changes was 108,701s, on average 253 ms per diff. 

Because we label each onion site by hand, we can check whether the type of site makes a difference in the annotation rate.
Table \ref{tab:success} shows an onion site's purpose against the success or failure of the annotation.
The annotation is most successful for the sites that promote a property offence, such as carding sites.
The annotation is least successful for sites promoting violence.
The difference is statistically significant ($p < 0.0001$).


\begin{table}
\caption{Crosstabulation of offense versus matching (10\% sample; N = 23,478,389; $p < 0.0001$)}
\label{tab:success}
\begin{tabular}{lrr|r}
\toprule
Offence vs match&Failure	&Success	&Total \\
\midrule
Child Sexual	& 73,217	& 101,234	& 174,451 \\
offence		& (42.0\%)	& (58.0\%)	& (0.7\%) \\
Drugs Offence	& 164,572	& 46,940	& 211,512 \\
		& (77.8\%)	& (22.2\%)	& (0.9\%) \\
No offence	& 502,821	& 255,167	& 757,988 \\
		& (66.3\%)	& (33.7\%)	& (3.2\%) \\
Property offence& 628,204	& 21,667,961	& 22,296,165 \\
		& ( 2.8\%)	& (97.2\%)	& (95.0\%) \\
Violent offence & 27,607	& 10,666	& 38,273 \\
		& (72.1\%)	& (27.9\%)	& (0.2\%) \\
\midrule
Total		&1,396,421	& 22,081,968	&23,478,389 \\
		& ( 6.0\%)	& (94.0\%)	& (100.0\%) \\
\bottomrule
\end{tabular}
\end{table}

Another way to evaluate the success of the annotations is to look at the most frequent annotation.
Table \ref{tab:annotations} lists the top ten annotations.
The last column gives an example of an insert/delete with context (in black) for each of the top ten annotations.
The explanation for the top ten is as follows:

(1) A Bitcoin address is a random-looking string recognised by the regular expressions if, for example, the text \\
https://www.blockchain.com/btc/address/ precedes it.

(2) A price must be preceded or followed by a currency symbol, such as BTC.

(3) Dates in standard formats can be recognised without context.

(4) {\bf fail} A random string is not recognised, even if the context, such as the word "wpnonce" here, suggests that it is a nonce.

(5) {\bf empty} is a technical tool to ensure that a sequence of changes of form c[{\color{red}d}{\color{green}i}c]+ can be represented as [{\color{red}d}{\color{green}i}c]+, where c stands for common text, {\color{red}d} for a delete and {\color{green}i} for an insert.

(6) Some onion sites advertise the number of users, transactions, orders, etc.

(7) An onion address is recognised if it is enclosed by "https?://" and ".onion".

(8) A time period in standard formats is recognised without context.

(9) An image filename should be followed by a standard image extension such as ".png" or ".jpeg".

(10) We assume that white space in HTML has no meaning.

The context is relatively specific in the above cases.
The list of regular expressions can also be expanded to cover other differences.


\begin{table}
\caption{The top ten annotations of deletes and inserts with examples (10\% sample; N = 23,478,389)}
\label{tab:annotations}
\begin{tabular}{lrrr}
\toprule
Annotation	& \%	& Examples \\
\midrule
Bitcoin		&47.7	&https://www.blockchain.com/btc/address/ \\
\multicolumn{3}{r}{{\color{orange}3Lc18ENW**********8mJt4zocpSAmFH}}\\
Price		&27.7	&{\color{red}0.00173}{\color{green}0.00174} BTC\\
Date		&14.5	&{\color{red}12/May}{\color{green}13/May}/2022\\
{\bf fail}	& 5.8	&wpnonce={\color{red}6d607}{\color{green}d08b4}\\
{\bf empty}	& 2.0	&\\
Ad		& 1.1	&{\color{red}17113}{\color{green}17132} Completed Orders\\
Space		& 0.3	& \\
Onion		& 0.3	&http://{\color{orange}facebookcorewwwi}.onion\\
Time		& 0.2	&{\color{red}8 days}{\color{green}2 weeks}\\
Image		& 0.1	&{\color{red}1-2}{\color{green}2-23}.png\\
\bottomrule
\end{tabular}
\end{table}

To validate the annotations, we checked a random sample of 453 annotations by hand and found one incorrect annotation.
This change is "BTC {\color{red}-1.90}{\color{green}1.18} \%", where the "BTC" suggests a {\color{blue}price} context, whereas a {\color{blue}percent} context is intended.

\paragraph{Step 5. Determine the chunk size}
Of the 443,194 homepage versions we analysed, 338,674 are unique (76.5\%). 
We can compare the number of unique homepage versions with the number of unique annotated templates, which is 101,349. 
\DarkDiff{} has characterised 70.1\% of the homepage versions as near-duplicates on the 10\% dataset.
We also know what causes the differences: Bitcoin addresses, prices, dates, advertisements, white space, onion addresses and time stamps (See Table \ref{tab:annotations}).

\begin{figure}[h]
\includegraphics[width=\linewidth]{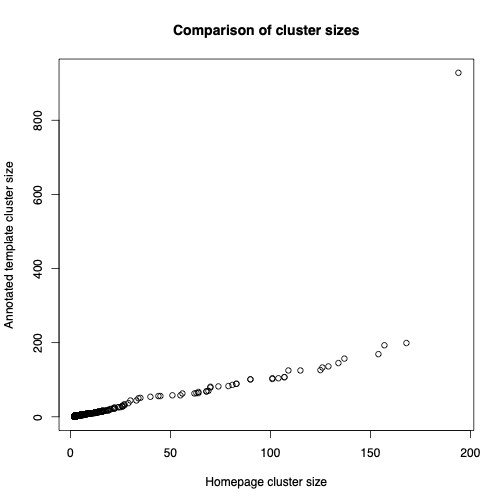}
\caption{Cluster sizes for homepages versus annotated templates}
\Description{Cluster size for original homepages versus annotated templates}
\label{fig:clusters}
\end{figure}

If two or more onion sites have an identical homepage or if they have identical annotated templates, we can cluster the sites.
Figure \ref{fig:clusters} shows a scatterplot of the number of onion sites per cluster for the homepages and the annotated templates.
There are 490 clusters for the homepages, with an average of 12.7 sites per cluster (median 4). 
For the annotated templates, there are 373 clusters with an average of 18.7 onion sites per cluster (median 5). 
The maximum cluster size of the annotated templates (928) is larger than the maximum for the homepages (194). 
The distribution of the cluster sizes is not normal, so we cannot investigate whether the mean cluster sizes differ significantly.
Instead, a non-parametric test (Wilcoxon rank sum test with continuity correction) shows that the difference in the median is significant ($p < 0.001$).

\subsection{Sensitivity analysis}
In the previous analysis, we chose chunk size ten.
Therefore, diffs are aligned in ten steps: first, the diffs of versions 0 \dots 9, then the diffs of versions 10 \dots 19 etc.
For the sensitivity analysis of the chunk size, we use seven different values: 0, 2, 5, 10, 20, 50, and 100.
The value 0 is the baseline because no alignment takes place.
Chunk size two corresponds to a minimum of 2x18=36 hours, over which the crawlers index versions and the largest chunk size 100 corresponds to 100x18 hours=75 days.
To limit the calculation time somewhat, we choose not 10\% but 1\% of the dataset for the sensitivity analysis.
We take the same 1\% sample for each chunk size value, which does not overlap with the 10\% sample.
After applying the exclusion criteria, 1,345 onion sites remain, with 43,248 homepage versions. 
Diff-match-patch takes an average of 162 ms per diff, comparable to the 132 ms for the 10\% sample. 
The annotation step takes an average of 341 ms per diff, comparable to the 253 ms for the 10\% sample. 
The number of unique pages for the 1\% sample is 37,976. 
The number of unique annotated templates varies with the chunk size as indicated in Table \ref{tab:sensitivity}.
The baseline of 44.9\% indicates that the alignment step produces a vital improvement. 
The best chunk size for the 1\% dataset is 10 or 20.
The percentage of near-duplicates with chunk size ten (68.4\%) is comparable to 70.1\% for the 10\% sample. 



\begin{table}
\caption{Sensitivity analysis of the chunk size (1\% sample; N=37,976)}
\label{tab:sensitivity}
\begin{tabular}{rrr}
\toprule
Chunk size	&Unique annotated&Near duplicates \\
		&templates	& \\
\midrule
baseline: 0 	&20,899		&44.9\% \\
  2		&14,090		&62.9\% \\
  5		&12,621		&66.8\% \\
 10		&11,987		&68.4\% \\
 20		&11,958		&68.5\% \\
 50		&12,396		&67.4\% \\
100		&13,858		&63.5\% \\
\bottomrule
\end{tabular}
\end{table}


\subsection{Rule maintenance}
Rules require maintenance to ensure they are up to date.
To show how much maintenance is needed, we (1) repeated the experiment with another 1\% sample of all data collected by the \DarkWebMonitor{} in the first five months of 2023 and (2) adjusted the rules.
The 2023 1\% sample contains 643 domains with 16,786 homepage versions, of which 14,432 are unique.
The rule changes consisted of adding the currency symbols for the Dollar and Ruble, commas in amounts such as 1,234.00 BTC, and words in the ad list, such as N "orders".
\DarkDiff{} generated 4,404 annotated templates for the 2023 1\% sample, resulting in a reduction of 69.5
This percentage is comparable to the reductions realised for the 10\% and the 1\% 2020-2022 samples.

\subsection{Comparison to Jaccard similarity}
The correlation of the Levenshtein similarity with the Jaccard similarity for 3-grams is 0.684, 5-grams 0.696 and 7-grams 0.700. 
All correlations are significant ($p < 0.0001$). 
Hence, the \DarkDiff{} similarity values correlate well with the similarities of locality-sensitive hashing schemes based on the Jaccard distance.

\subsection{Case Studies}
We take a closer look at three phenomena revealed by \DarkDiff{} while focusing on differences between homepage versions.

\paragraph{JavaScript}
We expect comparing pages with JavaScript to yield more differences than those without JavaScript.
Furthermore, the differences will be more difficult to recognise by regular expressions because the differences also often occur in the program text.
We, therefore, check to what extent JavaScript occurs in the \DWM{} dataset and what that means for the annotations.

Table \ref{tab:scripting} for the 10\% sample shows that the regular expressions for the annotations fail significantly more often ($p < 0.001$) on homepages with JavaScript than without.


\begin{table}
\caption{Crosstabulation of scripting versus matching (10\% sample; N = 23,478,389; $p < 0.0001$)}
\label{tab:scripting}
\begin{tabular}{lrr|r}
\toprule
Script vs match	& Failure 	& Success	&Total \\
\midrule
no scripting 	& 551,447	& 20,628,515	&21,179,962 \\
		& (2.6\%)	& (97.4\%)	& (90.2\%) \\
scripting 	& 844,974	& 1,453,453	& 2,298,427 \\
		& (36.8\%)	& (63.2\%)	& (9.8\%) \\
\midrule
Total 		&1,396,421	& 22,081,968	&23,478,389 \\
		& (6.0\%)	& (94.0\%)	& (100.0\%) \\
\bottomrule
\end{tabular}
\end{table}

\paragraph{Titles indicating service disruption}
The homepage is a website's business card, and the homepage's title summarises that.
That is why the differences between the titles of the successive homepage versions are interesting.
\DarkDiff{} found several differences in the home page's title, pointing at service disruption.
Suppose a website is under maintenance, which is helpful to know for visitors.
However, hackers would also be interested because the OPSEC of a regular website may be better than a site under maintenance.
We will investigate whether changes in the title of homepage versions indicate service disruption.

We analysed the titles of all versions of the homepage of all 393,291 onion sites in the 2020-2022 dataset. 
Suppose there are ten versions of one domein with the following titles [1,1,1,2,3,3,1,1,1,1]; then we group consecutive equal titles as follows: [1,2,3,1].
There are 895,201 such groups of titles in the entire dataset, on average 2.27 groups per domain.
Therefore, the homepage's title changes more than once during the period covered by the dataset (about two years). 
In addition to the regular homepages, we found the following types of homepages related to service disruption:
(1) "Alertmanager" and "Node Exporter" are pages generated by the Prometheus system that collects and provides information about websites.
(2) "404 NOT Found" and "Down for maintenance" are messages to the end user that the requested page is (temporarily) unavailable.
(3) "Welcome to Nginx!" and "Apache2 Ubuntu Default Page: It works" indicates that the web server has just been (re)installed, but no content is available yet.
(4) "Error \dots": A title with an error message indicates, for example, that the site is temporarily overloaded or that the database is unavailable.

We found 2,980 sites (0.8\%) which had a homepage title from the above list at some point in time. 
We found leaks of the server type (Apache, Nginx) and server versions.
A hacker can use that information to find vulnerabilities in the server from the CVE database and exploit the vulnerability to attack the site.
We do not know to what extent clear-web sites leak information during the service disruption, but we advise web admins to remain vigilant.

\paragraph{Hidden Bitcoins}
\DarkDiff{} has revealed that sometimes compliant onion addresses are changed into non-compliant ones.
A compliant V3 address must at least include 56 letters or numbers 2-7 and end with the letter 'd' \cite{Eaton2022}.
In the (blinded) example below, the red onion address is V3 compliant, but the green insert is not.
{\small "http://{\color{red}zhi 3q54n5a3fo**********fz735t2k3famqejsvht3 o7dpxevplceid}
{\color{green}zhi bc1q gzzf6un53f**********uk95jn92qlph4hecza o7dpxevplceid}.onion"}
The insert contains the beginning of a Bitcoin address {\bf bc1q}.
The beginning {\small "zhi"} and the ending {\small "o7dpxevplceid"} of both the delete and the insert are the same, but what is in between differs.
The part of the green insert {\small "{\color{green}bc1qgzzf6un53f**********uk95jn92qlph4hecza}"} is a valid Bitcoin address where transactions have taken place.
The corresponding part of the red delete {\small "{\color{red}3q54 \dots vht3}"} is not a valid Bitcoin address.

We analysed the oldest homepage of all 393,291 onion sites in the 2020-2022 dataset.
On every homepage, we searched all attributes of all HTML elements for hidden Bitcoin addresses, starting with "{\small bc1q}"
So we searched for hidden Bitcoin addresses in the onion addresses, but also in, for example, embedded images and JavaScript code.
We found 364 onion sites with 863 unique, valid Bitcoin addresses. 
Most were hidden in hyper links (55.2\%) and embedded images (12.1\%). 
Of the pages with hidden Bitcoin addresses, the crawlers discovered 99.9\% in 2022, the first 27.9\% in January 2022, and the last 0.2\% in September 2022. 
The Bitcoin addresses appear not only hidden but also visible to a visitor to the page.
A total of 0.18 BTC (approximately 3,300 Euro) has been deposited on 28 of the Bitcoin addresses found. 
There were yet to be any deposits on the other Bitcoin addresses.

In the example above, {\small "{\color{red}3q54 \dots vht3}"} has been mistaken for a Bitcoin address.
Bitcoin addresses can start with a 3, but this alleged address has 40 characters, while the maximum length is 35.
Probably all 364 onion sites with the same mistake are under the control of the same person, and some victims have sent some money to that person.

\section{Discussion}
We begin by answering the research question.
\DarkDiff{} uses Google's diff-match patch to calculate the differences between two homepage versions.
Most differences consist of specific dates, prices, Bitcoin addresses, etc.
Those differences can be recognised with regular expressions and replaced by a reserved word to create annotated templates.
We declare two web pages near-duplicates if the derived annotated templates are identical.
Therefore, \DarkDiff{} recognises near-duplicates, and the reserved words indicate the explanation.

As a dataset, we use random samples of homepages of onion sites to find and annotate differences between successive versions of the homepages.
We show that a limited number of regular expressions can recognise more than 90\% of all differences between subsequent homepage versions.
We then determined that about 70\% of the homepage versions are near-duplicates, and the reserved words explain why those versions differ.
This level of transparency cannot be achieved with locality-sensitive hashing (LSH) methods typically used to recognise near-duplicates.
\DarkDiff{} requires more computation time (0.35s per diff) than LSH methods.
The hash of the annotated templates can be stored in a database so that the efficiency of searching for previously discovered annotated templates is comparable to LSH methods.

The correlation of the Levenshtein similarity based on diffs with the Jaccard similarity is high.
Therefore, the difference between LSH-based techniques and \DarkDiff{} is not in the algorithm used (diffing vs Jaccard) but in the annotations of the differences with regular expressions.

Our results are obtained over a randomly selected subset of recent onion sites and their homepages from the \DWM{} dataset.
Our results are representative of the \DWM{} collection of recent onion sites.
To our knowledge, this is the most extensive collection of onion sites available for scientific research.
See Appendix \ref{app:benchmark} for more information.

\DarkDiff{} creates an annotated template where annotations such as {\color{blue}price} replace the inserts and deletes that change the previous version of a page to the current one.
Most of the \DWM{} dataset is financially motivated and regularly updated.
Therefore, most of the differences are Bitcoin addresses, prices and dates.

We separately analysed seven smaller random samples of the dataset with different alignment parameter values for a sensitivity analysis.
The percentages of near-duplicates improve with positive alignment vs the base case of no alignment.
For future work, we suggest automatically selecting the chunk size to optimise the success of the annotations.

We verified by hand the correctness of assigning reserved words via the regular expressions to changes and found one mistake in a sample of 450 annotations.
The mistake is due to ambiguity in natural language.

\DarkDiff{} highlights differences between homepages, and we found some notable differences that we investigated further.

(1) We compared pages with and without JavaScript, showing that JavaScript produces relatively more differences than can be recognised by regular expressions.
That is because program text is too complex for regular expressions to recognise.
\DarkDiff{} is more effective on the onion sites than on clear-web sites because JavaScript is used more often on the clear web.

(2) We found differences between onion addresses where valid Bitcoin addresses were hidden in the differences.
One or more scammers copied onion sites on a relatively large scale, attempting to replace existing Bitcoin addresses with their own.

(3) The crawlers of the \DarkWebMonitor{} regularly visit a page and store each unique version of that page.
If two subsequent versions show more differences than usual, the website could be more vulnerable to attack.
We have found about 2500 sites where the OPSEC of the onion site may be at stake.

\DarkDiff{} is file format agnostic; it works for any text file format that can be input to diff-match-patch.

\section{Limitations}
JavaScript is disabled in the crawlers of the \DarkWebMonitor{} to improve the performance of the crawler and to minimise security risks.
As a result, it is not possible to download and analyse dynamically generated content \cite{Nikiforakis2012}.

The \DarkWebMonitor{} has collected onion sites and pages for ten years.
There may be a bias in the crawler's sampling mechanism.

The regular expressions for the annotations have been designed to err on the side of caution, but false positives are possible.
The collection of regular expressions is easy to extend.

Detecting client-side redirects is a heuristic.

\section{Conclusions}
On the clear web, locality-sensitive hashing (LSH) is used to detect near-duplicates.
LSH is efficient but does not explain why two web pages are different.
\DarkDiff{} is an alternative to LSH that does provide such an explanation.
However, the explanations are encoded by regular expressions that need maintenance.

\DarkDiff{} focuses attention on the differences between web pages.
Most differences in the \DWM{} dataset can be explained automatically by a list of regular expressions.
However, differences that cannot be explained sometimes indicate interesting information about websites, for example, that they are under maintenance or that a scammer is at work.

\section*{Acknowledgements}
We thank Rolf van Wegberg for his comments on the paper.
The source code of the analysis programs is available from \url{https://doi.org/10.5281/zenodo.8050938}.
To access the data, please contact \url{https://www.cflw.com}.

\begin{wrapfigure}{r}{0.1\textwidth}
\includegraphics[width=\linewidth]{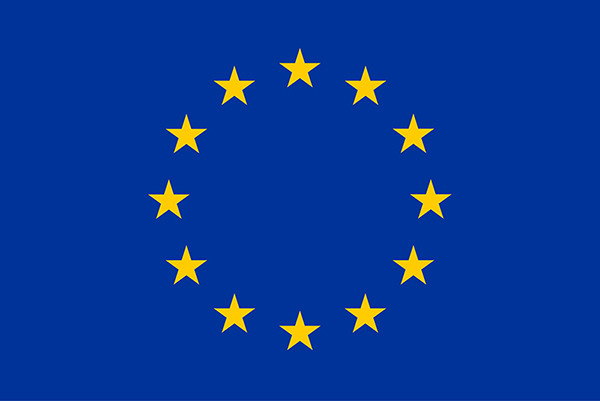} 
\end{wrapfigure}
The work described in this paper is performed in the H2020 project STARLIGHT (“Sustainable Autonomy and Resilience for LEAs using AI against High priority Threats”).
This project has received funding from the European Union's Horizon 2020 research and innovation program under grant agreement No 101021797.

\bibliographystyle{ACM-Reference-Format}
\bibliography{darkweb_refs.bib}

\appendix

\section{Comparison to a benchmark}
\label{app:benchmark}
To check how representative the \DWM{} dataset is for the number of existing onion sites, we collected a benchmark of onion sites from the literature and the clear web.
From the literature we have the Duta-10K dataset with 10K onions \cite{Al-Nabki2019}, the Ail dataset with 39K onions \cite{Falconieri2019}, and the Dizzy pages-basic dataset with 21K onions \cite{Boshmaf2023}.
The CoDa dataset with 13K onion \cite{Jin2022} does not contain onion addresses, nor their hash, so we could not include the CoDa dataset.
From the clear web, we use the blacklist from Ahmia.fi with 65K onion hashes and the DanWin dataset with 20K onions from onions.danwin1210.de.
We do not have access to the commercial datasets of darkowl.com and s2w.inc.

Because several datasets do not contain onion addresses but their MD5, we calculated the MD5 of all onions to compare the \DWM{} dataset with the benchmark.
The \DWM{} dataset, collected over the last 10 years, contains 3150K onions, 96\% of which are only in the \DWM{} dataset; 3.8\% appear in the \DWM{} dataset and the benchmark, and 0.2\% (4.5K onion addresses) appear only in the benchmark.
These are almost exclusively CSAM onion sites.
The conclusion is twofold.
On the one hand, the \DWM{} dataset is more extensive than the benchmark.
On the other hand, given that the Tor metrics project reports that there are about 750K onion sites live and that most onion sites live only a short time, the \DWM{} dataset is large but still incomplete.
The reason is that the current Tor V3 protocol was designed to hide this information \cite{Hoeller2021}.

\section{Comparison to the web of the past}
\label{app:past}
To gain insight into the dynamics of onion sites, we repeated related research \cite{Cho2000} on the \DWM{} dataset.
We analysed the average change interval and visible lifespan of the entire dataset.
The granularity of the analysis is one day, which reduces the dataset to 18,215,417 homepage versions. 
The comparison with the average change interval from the 1990s is shown in Table \ref{tab:dynamics}.
In both cases, the average change interval is U-shaped.
The visible lifespan has the same trend in both cases but rises more steeply in our case.


\begin{table}
\caption{Comparison of the average change interval and visible lifespan to the clear web of the 1990s (N = 18,215,417)}
\label{tab:dynamics}
\begin{tabular}{llrrrrrr}
\toprule
Statistic	&Exp.		&$\leq 1$&$> 1$	& $> 1$	& $>1$	& $> 4$	& sum \\
		&		&day	&day	& week	& month	& months& sum \\
\midrule
av. change	&\cite{Cho2000}	& 0.23	& 0.15	& 0.16	& 0.16	& 0.30	& 1.00 \\
interval	&Ours		& 0.31	& 0.12	& 0.12	& 0.24	& 0.20	& 1.00 \\
visible		&\cite{Cho2000}	& 0.00	& 0.07	& 0.16	& 0.22	& 0.55	& 1.00 \\
lifespan	&Ours		& 0.00	& 0.02	& 0.06	& 0.11	& 0.81	& 1.00 \\
\bottomrule
\end{tabular}
\end{table}

\begin{figure}[h]
\includegraphics[width=\linewidth]{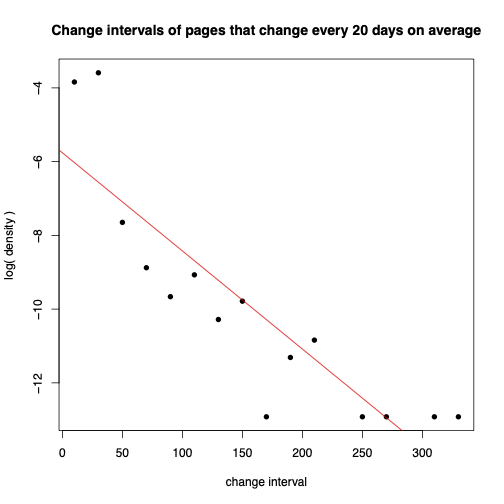}
\caption{Change intervals of pages (with 20-day average change interval)}
\Description{Change intervals of pages (with 20-day average change interval)}
\label{fig:intervals}
\end{figure}

We investigated whether the average visible lifespan of the homepage also follows a Poisson distribution.
The result in Figure \ref{fig:intervals} agrees with the comparable figure of Cho and Garcia-Molina \cite{Cho2000}.
The equation of the red line is log( average visible lifespan ) = -5.7 - 0.027 * average change interval. 
The adjusted $R^2$ is high (0.75), and the model and coefficients are statistically significant ($p < 0.0005$).

The dynamics of homepage versions of onion sites are comparable to the results of Cho and Garcia-Molina \cite{Cho2000}.
Their dataset stems from the 1990s, which suggests that onion sites resemble the clear web of the 1990s.
Onion sites use the same technical tools as modern clear-web sites but in different ways due to the difference in trust relationships.
For example, users of onion services have to make an effort to make sure they are not visiting a fake site.
Advertisers are also hesitant to pay for advertisements for the same reason.
As a result, web pages on onion sites are less dynamic than on the clear web.

\section{Primary data set and secondary analysis}
\label{app:secondary}
We describe the background and the collection of the primary information, which therefore falls outside the scope of the \DarkDiff{} investigation, but is relevant for the ethical aspects of the secondary analysis.

The primary dataset was collected to contribute to a safer society by analyzing services on the Dark Web. These are often criminal in nature and are used as intelligence or evidence by parties working towards a safer society. By being transparent about the \DarkDiff{} investigation, we hope to contribute to the transparency of the tools law enforcement uses and the transparancy of the judiciary. \DWM{} uses a standard snowballing process to fetch web pages from the darknet. The process starts with a set of start addresses that are downloaded. New addresses found on the downloaded pages are also retrieved, etc. This process is also used on the clear web by others, such as Google.

We now follow the Menlo report \cite{Dittrich2012} to discuss the ethical risks (R) and mitigations (M).

\subsection {Respect for Persons}
R: Most dark websites only contain data that is difficult to trace back to a person, such as onion addresses and Bitcoin addresses.
M: In the secondary analysis, we replaced as much information from web pages as possible with reserved words such as “Bitcoin”, “Price”, and “Date”.

R: Although unlikely, it is possible that a malicious and a benign website have the same annotated template and are in the same cluster.
M: We have not published any onion addresses and only report on the size of the clusters.

R: In the secondary analysis, we only looked up Bitcoin addresses on the public blockchain that we thought were involved in fraudulent acts.
This allowed us to report the size of the fraudulent acts (in Euro).
M: We did not in any way try to find out the owner of an Onion site or a Bitcoin address.

\subsection{Benefit}
R: Law enforcement may approach owners of malicious websites sooner than they would without using \DarkDiff{}, harming those owners.
M: According to the principle of proportionality, we argue that the benefits to law enforcement (and society) outweigh the harms to the owners of malicious onion sites.

R: The owner of malicious onion sites who learns about how \DarkDiff{} works could take advantage of this by making the annotations more difficult.
M: We believe that the advantage of transparency about the research outweighs the disadvantage that owners of malicious onion sites use the publication.

\subsection{Justice}
R: Because the size of the Dark Web is unknown and cannot be known \DarkDiff{} research may be biased.
M: As far as we can tell, the \DWM{} dataset is one of the largest collections, which was collected without known bias.

\subsection{Respect for Law and Public Interest}
R: The Darknet is used to distribute CSAM.
M: \DarkDiff{} only performs textual analysis, and as an extra precaution, all CSAM pages with embedded images are excluded from secondary analysis.

R: The \DarkDiff{} study showed that 97\% of the data is related to crime, which puts the researchers at risk.
M: During the research, we regularly reported our findings to law enforcement, who took or could have taken timely action.

R: The working methods of the customers of \CFLW{} must remain secret.
M: \DarkDiff{} is one example of the many tools the company makes available to its customers, and knowing how \DarkDiff{} works says nothing about how the customers work.

R: The clients of \CFLW{} may add onion addresses to the primary data in an ongoing investigation.
M: Information about ongoing investigations is confidential and carefully excluded from secondary analysis.

R: Is it possible that the research has overburdened the TOR network?
M: The \DarkDiff{} research has not yielded any traffic on the TOR network because it is secondary analysis.

R: Did \CFLW{} run any additional risks from the \DarkDiff{} investigation?
M: The secondary analysis was done entirely on servers of \CFLW{}, and all data remained within the intranet.

\end{document}